\documentclass[aps,twocolumn,showpacs,superscriptaddress,amsmath,amssymb,pre,nofootinbib]{revtex4-1}
\usepackage{epsf,amsmath,amssymb,amsfonts,verbatim,color,multirow,pifont}
\usepackage{graphicx}
\usepackage{dcolumn}
\usepackage{bm}
\usepackage{txfonts}
\usepackage{hyperref}
\usepackage{soul}

\begin{document}

\title{Thermodynamic cost of synchronizing a population of beating cilia}

\author{Hyunsuk Hong}
\affiliation{Department of Physics and Research Institute of Physics and Chemistry, Jeonbuk National University, Jeonju 54896, Korea}

\author{Junghyo Jo}
\affiliation{Department of Physics Education, Seoul National University, Seoul 08826, Korea}

\author{Changbong Hyeon}
\email{hyeoncb@kias.re.kr}
\affiliation{School of Computational Sciences, Korea Institute for Advanced Study, Seoul 02455, Korea}

\author{Hyunggyu Park}
\email{hgpark@kias.re.kr}
\affiliation{School of Physics, Korea Institute for Advanced Study, Seoul 02455, Korea}

\date{\today}

\begin{abstract}
Synchronization among arrays of beating cilia
is one of the emergent phenomena in biological processes at meso-scopic scales.
Strong inter-ciliary couplings modify the natural beating frequencies, $\omega$, of individual cilia to produce a collective motion that moves around a group frequency $\omega_m$.
Here we study the thermodynamic cost of synchronizing cilia arrays
by mapping their dynamics onto a generic phase oscillator model.
The model suggests that upon synchronization the mean heat dissipation rate is decomposed into two contributions, dissipation from each cilium's own natural driving force and dissipation arising from the interaction with other cilia, the latter of which can be interpreted as the one produced by a potential with a time-dependent protocol in the framework of our model.
The spontaneous phase-synchronization of beating dynamics of cilia induced by strong inter-ciliary coupling is always accompanied with a significant reduction of dissipation for the cilia population, suggesting that organisms as a whole expend less energy by attaining a temporal order.
At the level of individual cilia, however, a population of cilia with $|\omega|< \omega_m$ expend more amount of energy upon synchronization.
\end{abstract}

\maketitle

\section{Introduction}
Spatiotemporal dynamics and pattern formation that emerge in living organisms have been an abiding interest in biological physics for many decades~\cite{winfree2001book,cross1993RMP}.
Metachronal coordination in arrays of beating cilia that cover the surface of various organisms is one of the striking examples that highlight the synchronous interactions of biological organisms at the cellular level.
Although inter-ciliary mechanochemical feedback control is still considered as a possible mechanism for collective dynamics,
there has been a longstanding hypothesis as well as experimental demonstrations that hydrodynamic inter-ciliary coupling alone is sufficient to produce synchronous dynamics in low Reynolds number environments~\cite{Purcell77AJP,gueron1997PNAS,kim2006PRL,guirao2007BJ,goldstein2009PRL,Elgeti2013PNAS,sanchez2011Science}.

Cellular environment is replete with free energy sources maintained via homeostasis \cite{AlbertsBook}, and thus the energy itself may not be the main concern for individual cellular processes.
However, when both the energy-consumption rate and the number of such energy-consuming components are increased,
the biological system as a whole would soon confront a shortage of energy supply.
In such a case, reducing the total amount of energy consumption would become the key priority.
In fact, in his seminal paper~\cite{Taylor1951RoyalSoc}, G. I. Taylor analyzed the dynamics of a pair of fluctuating sheets to suggest that in-phase coordination of two sheets is more advantageous because it reduces the rate of \emph{energy dissipation}, the value of which was calculated in terms of the work done by the sheet against viscous stress.
Such consideration was later followed up by many researchers~\cite{gueron1999PNAS,mettot2011PRE}.

\begin{figure}[t]
        \includegraphics[width=1.0\linewidth]{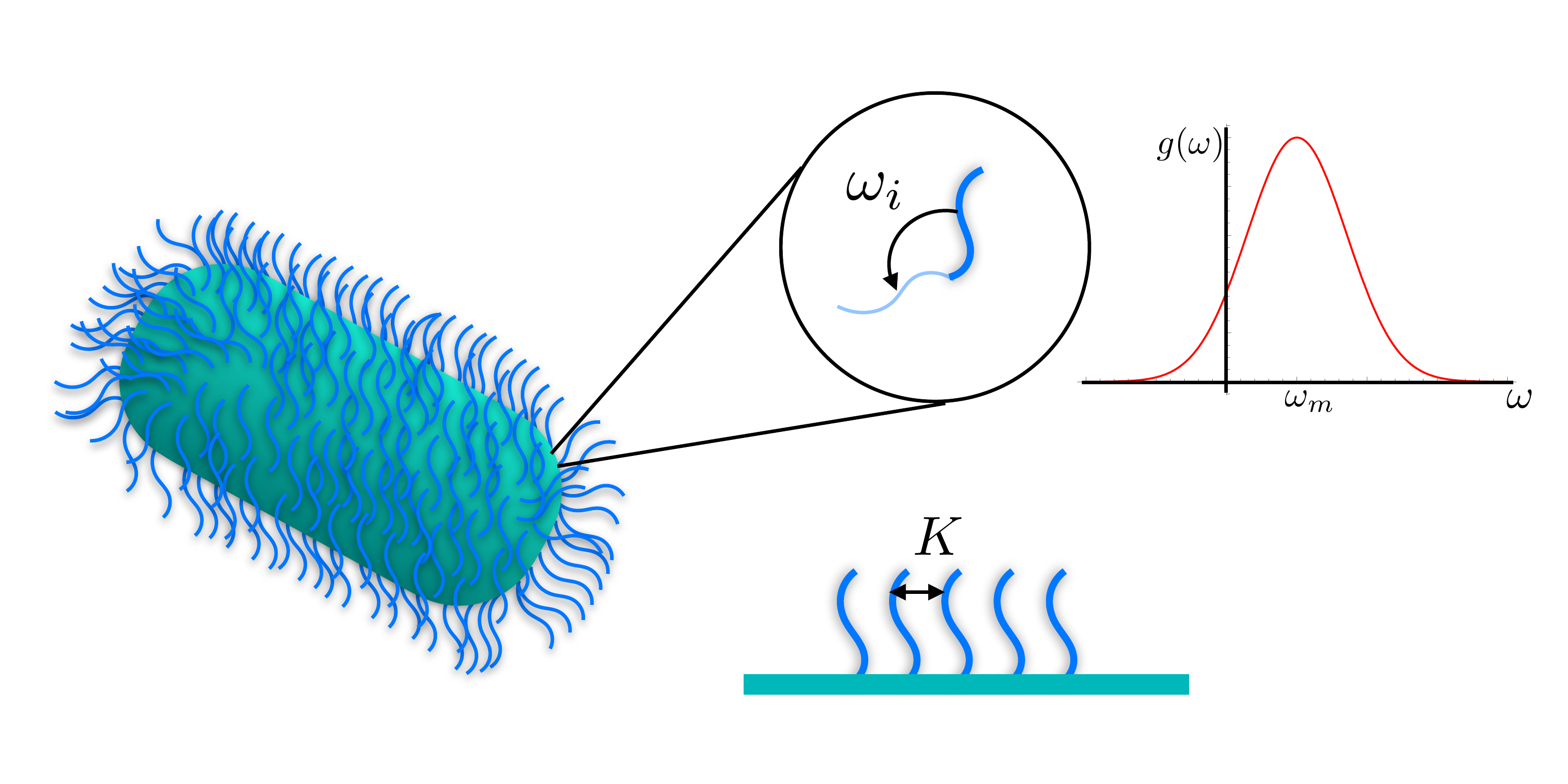}
        \caption{\label{fig:model}
        Illustrated are the cilia covering the surface of a bacterium.
        The periodic beating dynamics of individual cilium is driven by its intrinsic frequency $\omega_i$.
        The frequency could differ from one cilium to another.
        To model such a heterogeneity in cilia dynamics, $\omega_i$ for the $i$th cilium is selected from a distribution $g(\omega)$ whose mean value is $\omega_m$.
        The metachronal wave on the bacterium surface could emanate from the synchronization of beating cilia. The strength of the hydrodynamic coupling between cilia is given by $K$.
}
\end{figure}

Here we extend the foregoing energetic consideration to a statistical mechanical level by mapping a set of coupled arrays of cilia onto a noise-dressed version of the Kuramoto oscillator model \cite{kuramoto2003book}, where the phase dynamics of an individual oscillator is described by a following set of coupled equations~\cite{sakaguchi1988PTP,SonHong10}:
\begin{equation}
\dot\phi_i = \omega_i - \frac{K}{N}\sum_{j=1}^N\sin(\phi_i-\phi_j)+\eta_i(t)
\label{eq:model1}
\end{equation}
with $i=1, 2,\dots, N$. The phase variable $\phi_i$ represents the beating motion of the $i$-th cilium characterized with its own natural driving frequency $\omega_i$.
With an assumption of \emph{cilium-to-cilium heterogeneity}, which is supported by experimental observations \cite{dey2018PRE,aubusson2015ciliary},
the natural frequency $\omega_i$ in the first term
could be chosen from a distribution function, $g(\omega)$.
In this study, we consider a Gaussian function,
$g(\omega_i)=\mathcal{N}(\omega_m,\sigma^2)$, with the mean $\omega_m$ and variance $\sigma^2$, as a model of heterogeneous cilia population.

In the second term, provided that the hydrodynamics is the origin of the inter-ciliary coupling,
the parameter $K(>0)$ should be a function of the cilium length and the inter-cilia distance with its strength depending on the geometrical detail of a pair of cilia.
For simplicity, however, we set $K$ constant for any cilia pair, assuming a mean-field type all-to-all coupling.
The phase difference between $\phi_i$ and $\phi_j$ is minimized for large $K$, giving rise to the cilia's coordinated beating motion.

The last term $\eta_i(t)$, which is essential for calculating the heat dissipated from the system,
is modeled using the Gaussian noise that satisfies $\langle\eta_i(t)\rangle=0$ and
$\langle \eta_i(t) \eta_j(t') \rangle = 2D\delta_{ij}\delta(t-t')$.
The noise represents the ambient thermal environment with temperature $T$, surrounding the cell with cilia and its strength obeys the Einstein relation, $D=k_B T/\gamma$, where $\gamma$ is a friction coefficient of each cilium. We set $\gamma=1$ for convenience throughout
this paper.

In this work, we quantify the thermodynamic cost (or heat dissipation) for a population of beating cilia upon synchronization, which is modeled with Eq.~\eqref{eq:model1}.
As is well studied in the past, for the entire cilia population, the total mean dissipation is reduced upon synchronization for a $K$ value greater than its threshold $K_c$.
Our careful analysis, however, discovers that the mean dissipation from a single cilium upon synchronization with others can be greater than in isolation if its natural frequency ($\omega_i$) is smaller than the average frequency of the population $\omega_m\equiv (1/N)\sum_{i=1}^N\omega_i$.

In Sec.~II, the mean-field version of noisy Kuramoto model is introduced to describe the interacting cilia and their synchronization.
In Sec.~III, we calculate the mean heat dissipation from individual cilia as well as from the entire cilia population in disordered and synchronized phases.
A special attention will be paid to a physically correct way of calculating the heat dissipation to comply with the 2nd law of thermodynamics.
Finally, we conclude with the significance of our work in light of the thermodynamics of many-body synchronization.

\section{Noisy Kuramoto model}

The equation of motion for the noisy Kuramoto model, Eq.~\eqref{eq:model1}, can be cast into a simple form
\begin{align}
\dot\phi_i = \omega_i - Kr\sin{[\phi_i-\theta(t)]}+\eta_i(t)~,
\label{eq:Sakaguchi}
\end{align}
with the synchronization order parameter defined as~\cite{kuramoto2003book}
\begin{equation}
r e^{i\theta} \equiv \frac{1}{N}\sum_{j=1}^N e^{i\phi_j}
\label{eq:r_and_theta1}
\end{equation}
where  the order parameter $r$ measures the extent of phase coherence ($0\leq r \leq 1$), and
$\theta$ is the average phase angle.
It is well known that the system reaches a steady state in the long-time limit,
where $r$ with $\langle\delta r^2\rangle\sim1/{N}$ becomes a time-independent constant in the $N\rightarrow\infty$  limit.
Also in this limit, the average angle $\theta$ varies linearly in time as
$\theta(t)=\omega_m t$ with a mean velocity (or a group velocity) defined as
\begin{align}
\omega_m\equiv \frac{1}{N}\sum_i^N\omega_i.
\end{align}
In this paper, we confine ourselves to the steady state behavior of cilia population.

It is more convenient to rewrite the equation of motion with a {\em shifted} phase variable $\tilde\phi_i$ as
\begin{align}
\dot{\tilde\phi}_i=\tilde\omega_i - Kr\sin{\tilde\phi_i}+\eta_i(t)~
\label{eq:Sakaguchi_s}
\end{align}
with ${\tilde\phi}_i\equiv \phi_i-\theta(t)$ and $\tilde\omega_i\equiv \omega_i-\omega_m$.
The distribution
for the shifted natural frequency $\tilde\omega_i$ becomes symmetric Gaussian, i.e.~${\tilde g}(\tilde\omega_i)=\mathcal{N}(0,\sigma^2)$.
As seen in Eq.~\eqref{eq:Sakaguchi_s},
all oscillators (cilia) become independent to each other with fixed $r$, and then the probability distribution
function (PDF) of the total system $\rho_\textrm{tot}$ is simply the product of the PDF of each oscillator as
\begin{align}
\rho_\textrm{tot}=\prod_{i=1}^N \rho_i (\tilde\phi_i)~.
\end{align}

In the steady state, each PDF $\rho_i$ can be calculated exactly in the standard Fokker-Planck framework as~\cite{sakaguchi1988PTP,SonHong10}
\begin{align}
\rho(\tilde\phi,\tilde\omega)= \frac{e^{-V(\tilde\phi,\tilde\omega)/D}}{Z(\tilde\omega)}
\Bigg(1-\frac{1-e^{-2\pi\tilde\omega/D}}{\int_0^{2\pi}d\phi'e^{V(\phi',\tilde\omega)/D}}\int_0^{\tilde\phi}d\phi''e^{V(\phi'',\tilde\omega)/D}\Bigg)~,
\label{eq:pdf_s}
\end{align}
where the subscripts `$i$' are dropped for $\tilde\phi$ and $\tilde\omega$ for simplicity
and the {\em potential} function
$V(\tilde\phi,\tilde\omega) \equiv -\tilde\omega\tilde\phi-Kr\cos(\tilde\phi)$.
Note that the PDF is a periodic function of phase, i.e.~$\rho(\tilde\phi,\tilde\omega)=\rho(\tilde\phi+2\pi,\tilde\omega)$ and the normalization constant
$Z(\tilde\omega)$ is determined by $\int_{0}^{2\pi} d\tilde\phi ~\rho(\tilde\phi,\tilde\omega)=1$ (see the explicit expression for $Z$ in
Eq.~\eqref{eq:app_z} of the Appendix A).

The order parameter equation \eqref{eq:r_and_theta1} becomes
\begin{align}
r  \equiv \frac{1}{N}\sum_{j=1}^N e^{i\tilde\phi_j}~,
\label{eq:r_s}
\end{align}
which provides a {\em self-consistency} relation to determine the value of $r$ in the steady state as
\begin{align}
r = \int_{-\infty}^{\infty} d\tilde\omega~ \tilde g(\tilde\omega) \int_{0}^{2\pi} d\tilde\phi~ e^{i\tilde\phi}
\rho(\tilde\phi,\tilde\omega)~,
\label{eq:r_byFP1}
\end{align}
where $r$ should be understood as the steady-state ensemble average of Eq.~\eqref{eq:r_s} in the
$N\rightarrow \infty$ limit.
Although the closed form of $r$ 
is not known,
the critical behavior of $r$ near the transition is obtained using a perturbation expansion
for small $r$~\cite{sakaguchi1988PTP,SonHong10}.
\begin{equation}
r
\sim (K-K_c)^{1/2}~\qquad \textrm{for}~~K\ge K_c~,
\label{eq:r_sc}
\end{equation}
where the threshold value $K_c$ for the transition is given by
\begin{align}
K_c=2 \left[\int_{-\infty}^{\infty} d\tilde\omega  \frac{D\tilde g(\tilde\omega)}{D^2 + {\tilde\omega}^2}\right]^{-1}~,
\label{eq:Kc}
\end{align}
and $r=0$ for $K\le K_c$. The steady-state PDF $\rho(\tilde\phi,\tilde\omega)$
can be calculated from Eq.~\eqref{eq:pdf_s} with $r$ obtained from Eq.~\eqref{eq:r_byFP1}. For $K\le K_c$, no
synchronization occurs, thus the PDF is uniform as $\rho=1/(2\pi)$.
\begin{figure*}[t]
        \includegraphics[width=0.8\linewidth]{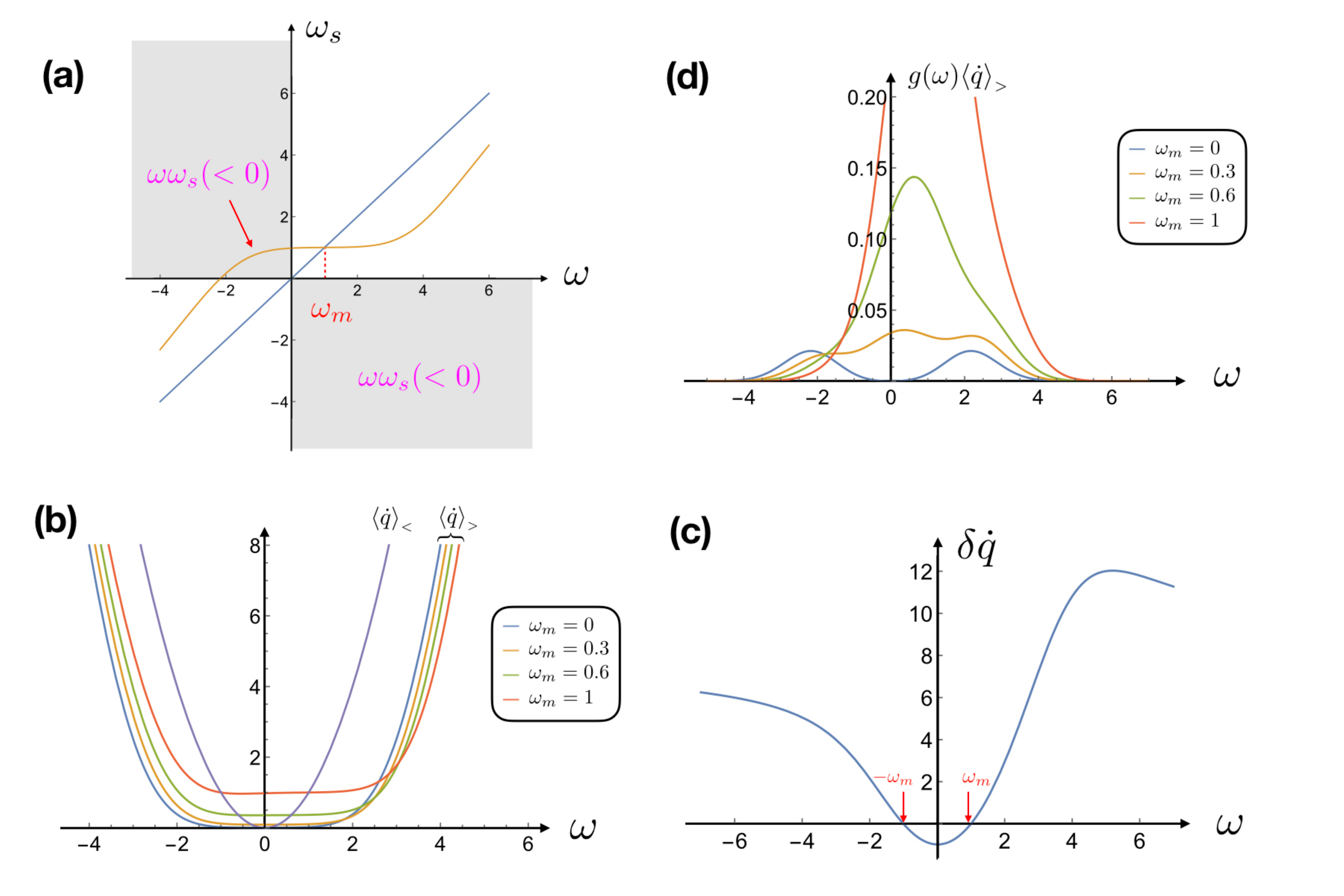}
        \caption{\label{fig:q}
        Quantities involving the mean heat dissipation rate calculated at $D=1$, $\sigma=1$, and $K=4>K_c$.
        The threshold value is given by $K_c=2(2/e\pi)^{1/2}/\text{erfc}(1/\sqrt{2})\approx 3.05$ from Eq.~\eqref{eq:Kc}.
        (a) Plots of the coupling modified frequency $\omega_s$ as a function of $\omega$ with $\omega_m=1$.
        The solid line in orange shows the $\omega$-dependence of $\omega_s$ for $K=4>K_c$ and the blue line 
        denotes $\omega_s=\omega$ valid for $K<K_c$.
The region of $\omega_s\omega<0$ is under the grey shadow.
The red arrow highlights the range of $\omega$ giving rise to the negative viscous dissipation ($\omega_s\omega<0$).
(b) Mean heat dissipation rate $\langle\dot{q}\rangle$ from individual cilia as a function of $\omega$.
For $K<K_c$, $\langle\dot{q}\rangle_<=\omega^2$.
For $K>K_c$, $\langle\dot{q}\rangle_>=\omega\omega_s+\omega_m(\omega_s-\omega)$, which is plotted for various $\omega_m$ values.
(c) Heat reduction rate upon synchronization, $\delta\dot{q}[\equiv\langle\dot{q}\rangle_<-\langle\dot{q}\rangle_>]$, versus $\omega$ with $\omega_m=1$.
(d) Population-weighted mean heat dissipation rate $g(\omega)\langle\dot{q}\rangle_>$ 
for various values of $\omega_m$.
}
\end{figure*}

It is useful to investigate the {\em coupling-modified} frequency $\tilde\omega_s$ of each oscillator, which is defined as
\begin{align}
\tilde\omega_s\equiv \langle \dot{\tilde\phi}\rangle =\langle\tilde\omega -Kr \sin {\tilde\phi}\rangle~,
\label{eq:omega_s}
\end{align}
where $\langle\cdots\rangle$ is the steady-state ensemble average. It is quite simple to calculate $\tilde\omega_s$ exactly, resulting in
\begin{align}
\tilde\omega_s  = \frac{2\pi D}{Z(\tilde\omega)}\frac{1-e^{-2\pi\tilde\omega/D}}{\int_0^{2\pi}d\phi'e^{V(\phi',\tilde\omega)/D}}
\equiv \tilde\omega\left[1-\alpha\right],
\label{eq:tilde_omega0}
\end{align}
where the modification factor $\alpha$ can be expressed as
\begin{align}
\frac{1}{1-\alpha}=
I_0^2(a) + 2\sum_{n=1}^{\infty}
 \frac{b^2(-1)^n  I_n^2(a)}{n^2 + b^2}~,
\label{eq:tilde_omega}
\end{align}
where $a\equiv Kr/D$, $b\equiv \tilde\omega/D$,  and  $I_n$ is the $n$-th order modified Bessel function of the 1st kind
(see the detailed derivation in the Appendix B). The modification factor $\alpha=\alpha(a,b^2)$, which ranges between 0 and 1, monotonically increases with $a$ (proportional to the synchronization order parameter $r$) and decreases with  $b^2$ (proportional to the square of the shifted frequency ${\tilde\omega}^2$).
At $r=0$ ($K\le K_c$), no modification
occurs ($\alpha=0$) with $\tilde\omega_s=\tilde\omega$, simply from Eq.~\eqref{eq:omega_s}. For $K\ge K_c$ (synchronized phase), oscillators should slow down
due to the coupling ($0\leq \alpha<1$), so $|\tilde\omega_s|<|\tilde\omega|$ for all $\tilde\omega$, and $\tilde\omega_s$ approaches  $\tilde\omega$ for large $|\tilde\omega|$. 

The coupling-modified frequency $\omega_s$ in terms of the original variables $\phi$ is given by $\omega_s=\langle\dot\phi\rangle=\tilde\omega_s+\omega_m$
with the natural frequency $\omega=\tilde\omega+\omega_m$. We plot $\omega_s$ against $\omega$ in Fig.~\ref{fig:q} (a).
Note that $\omega_s$ is close to the group velocity $\omega_m$  ($\tilde\omega\approx 0$) in the range of $|\omega-\omega_m|\lesssim Kr$ ($b\lesssim a$). In fact, we can easily find $1-\alpha\approx 2\pi a e^{-2a}$ for $b/a << 1$ from Eq.~\eqref{eq:tilde_omega}, leading to
\begin{align}
\omega_s\approx \omega_m+ \frac{2\pi Kr}{D}e^{-2Kr/D}(\omega-\omega_m) ~~\textrm{for}~~
|\omega-\omega_m|\lesssim Kr,
\end{align}
where the correction term is exponentially small for large $Kr/D$. In the other limit for large $|\omega|$ or small $r$ ($b>>a$), the modification factor $\alpha$ is negligible as
\begin{align}
\alpha\approx \frac{a^2}{2(1+b^2)}~\quad\textrm{for}\quad \frac{a}{b}<<1~,\label{eq:alpha_small}
\end{align}
and thus 
\begin{align}
\omega_s\approx \omega- \frac{1}{2}\frac{(Kr)^2(\omega-\omega_m)}{D^2+(\omega-\omega_m)^2}~\quad\textrm{for}\quad 
|\omega-\omega_m|>> Kr~,\label{eq:omega_small}
\end{align}
which approaches $\omega_s=\omega$ in the $r\rightarrow 0$ limit.

\section{mean dissipation from individual cilia}

The heat dissipation rate from a single cilium with a natural frequency $\omega$  is calculated as~\cite{sekimoto2010book}
\begin{equation}
\dot{q}(\omega)  =  \left[\omega-Kr\sin(\phi-\theta)\right]\circ \dot{\phi}(t),
\label{eq:q_i}
\end{equation}
which is the energy loss caused by the thermal force $F (\omega)=\omega-Kr\sin(\phi-\theta)$ in Eq.~\eqref{eq:Sakaguchi}
and the symbol $\circ$ denotes the Stratonovich multiplication~\cite{sekimoto2010book}.
The corresponding rate of work done on the cilium is
\begin{equation}
\dot{w} =  \omega \dot{\phi}(t)-\dot\theta Kr \sin (\phi-\theta)~,
\label{eq:w_i}
\end{equation}
where the first term is the rate of work done by the driving force $\omega$ and the second one is due to the \emph{Jarzynski work} rate,
$\dot\theta {\partial E}/{\partial\theta}$~\cite{JarzynskiPRL97},
associated with the time-dependent protocol $\theta(t)$ in the potential energy function $E(\phi,\theta)
=-Kr\cos (\phi-\theta)$.
Together with Eq.~\eqref{eq:q_i}, the thermodynamic first law for each cilium, $\dot{w}=\dot{E}+\dot{q}$,
with $\dot{E}(\phi,\theta)=(\partial_{\phi}E)\dot{\phi}+(\partial_{\theta}E)\dot{\theta}$ yields Eq.~\eqref{eq:w_i}.

In the steady state, $\langle \dot{E}\rangle=0$, thus the mean values of the heat dissipation and work production rates should be identical.
Then we get the mean heat dissipation rate as 
\begin{align}
\langle \dot{q}\rangle &= \langle \dot{w}\rangle \nonumber\\
 &= \omega \langle \dot{\phi}\rangle -\omega_m \langle Kr \sin (\phi-\theta)\rangle\nonumber\\
 & = \omega \langle \dot{\phi}\rangle + \omega_m (\langle\dot{\phi}\rangle -\omega),\nonumber\\
 &=\omega\omega_s+\underbrace{\omega_m(\omega_s-\omega)}_{\text{Jarzynski work}}~,\label{eq:w_i_av}
\end{align}
where 
$\omega_s(=\tilde{\omega}_s+\omega_m)$ can be obtained from 
Eqs.~\eqref{eq:tilde_omega0} and \eqref{eq:tilde_omega}.
Of particular note is that the dissipation due to $\omega\omega_s$ alone can be negative for a small subpopulation of cilia with $\omega<0$ and $\omega_s>0$ when $\omega_m>0$ (see the shaded region of Fig.~\ref{fig:q} (a)).
This negative dissipation due to $\omega\omega_s<0$, however, is compensated by the contribution from the Jarzynski work, giving rise to $\langle\dot{q}\rangle\geq 0$.
The heat dissipation $\langle \dot{q}\rangle$ ought to be always non-negative for all $\omega$ to be consistent with the
2nd law of thermodynamics (see Fig.~\ref{fig:q} (b)).

In the disordered phase ($K<K_c$), no modification is made to the natural frequency ($\omega_s=\omega$), and hence $\langle \dot{q}\rangle=\langle\dot{q}\rangle_<\equiv\omega^2$, where the subscript $<$ stands for the condition $K<K_c$.
It is, however, interesting to note that, for $\omega_m\neq 0$, the heat dissipation from a single cilium is not always reduced upon synchronization ($K>K_c$); instead, its sign is decided by the value of $\omega$ (see Figs.~\ref{fig:q} (b) and (c)).
With Eq.~\eqref{eq:w_i_av} and $\alpha(\omega)=(\omega-\omega_s)/(\omega-\omega_m)$ from Eq.~\eqref{eq:tilde_omega0}, it is easy to show that
\begin{align}
\delta\dot{q}&\equiv \langle \dot{q}\rangle_< -
\langle \dot{q}\rangle_>\nonumber\\
&=(\omega-\omega_s)(\omega+\omega_m)\nonumber\\
&=\alpha(\omega)(\omega^2-\omega_m^2).
\end{align}
Interestingly, for the cilia with the natural frequency in the range of $|\omega|\leq\omega_m$,
we obtain $\langle\dot{q}\rangle_>\geq \langle\dot{q}\rangle_<$ (Fig.~\ref{fig:q} (b)) or
$\delta\dot{q}\leq 0$ (Fig.~\ref{fig:q} (c)), suggesting that
more amount of heat is dissipated upon synchronization than the case in the disordered phase.

\begin{figure}[t]
        \includegraphics[width=1\linewidth]{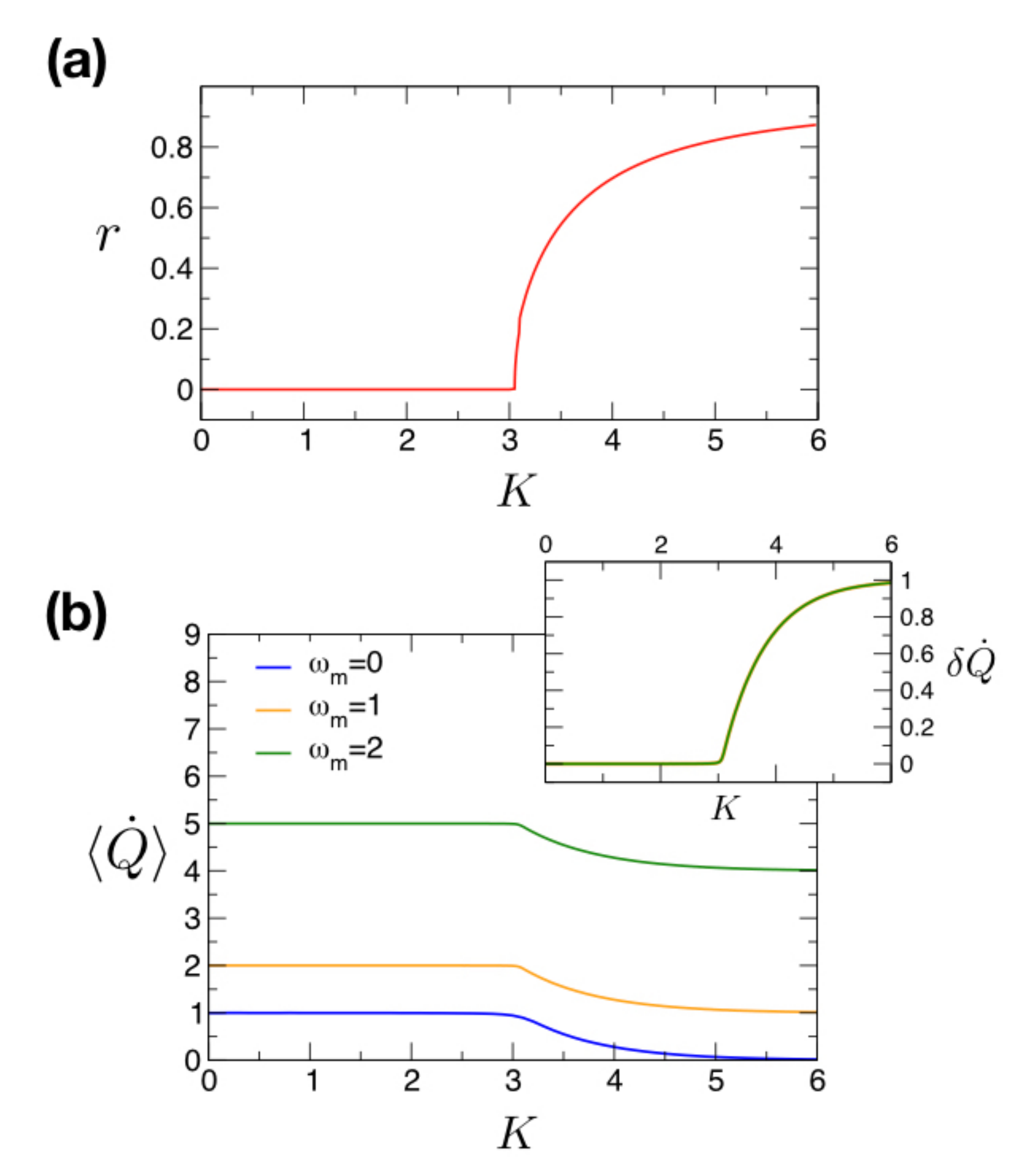}
        \caption{\label{fig:rQK}
        (a) The synchronization order parameter $r$ as a function of $K$.
        (b) The rate of total mean heat dissipation rate $\langle\dot{Q}\rangle$ as a function of $K$ with $\omega_m=0$, $1$, $2$, and $\sigma=1$.
        The inset shows $\delta\dot{Q}$ for $\omega_m=0$, $1$, and $2$.
        In consistent with the discussion in the main text, all the $\delta\dot{Q}$ for different $\omega_m$ collapse onto a single curve.
}
\end{figure}

The rate of total mean heat dissipation $\dot{Q}$ for the ensemble of $N$ cilia is obtained by integrating the 
population-weighted $\langle\dot{q}\rangle$ (see Fig.~\ref{fig:q} (d)):
\begin{equation}
\langle\dot{Q}\rangle \equiv \frac{1}{N}\sum_{i=1}^N \langle \dot{q}\rangle(\omega_i) =\int^{\infty}_{-\infty}d\omega~ g(\omega )\langle\dot{q}\rangle~.
\label{eq:Q}
\end{equation}
For $K<K_c$, $\langle\dot{Q}\rangle=\langle\dot{Q}\rangle_<\equiv \int^{\infty}_{-\infty} d\tilde\omega~ \tilde g(\tilde\omega) (\tilde\omega+\omega_m)^2=
\sigma^2+ \omega_m^2$. The total mean heat dissipation rate upon synchronization, $\langle\dot{Q}\rangle_{>}$, for $K>K_c$ is always smaller than $\langle\dot{Q}\rangle_{<}$ and is reduced by
\begin{align}
\delta\dot{Q}&\equiv \langle\dot{Q}\rangle_< -\langle\dot{Q}\rangle_> \nonumber\\
&=\int d\omega g(\omega)\left(\langle\dot{q}\rangle_<-\langle\dot{q}\rangle_>\right)\nonumber\\
&=\int d\tilde\omega~ \tilde g(\tilde\omega) {\tilde\omega}^2
\alpha(a,b^2) \ge 0 ~.
\label{eq:dQ}
\end{align}
Note that the amount of the total heat reduction $\delta\dot{Q}$ is independent of the frequency shift $\omega_m$ because $\alpha$ is a function of the shifted frequency $\tilde\omega$ only ($b=\tilde\omega/D$). The frequency shift $\omega_m$ only comes in as
a simple constant addition of $\omega_m^2$ to the total heat dissipation.
In addition, unlike $\delta\dot{q}$ (Fig.~\ref{fig:q} (c)), $\delta\dot{Q}$ is always positive and universal for any value of $\omega_m$ (see the inset of Fig.~\ref{fig:rQK} (b)), signifying that the thermodynamic cost of the entire cilia population is always reduced upon synchronization.

Finally, the nature of the synchronization transition at small $r$ in the vicinity of $K\approx K_c$ is of interest.
Using Eq.~\eqref{eq:alpha_small} for small $r$, the total heat reduction rate becomes 
\begin{align}
\delta\dot{Q}\approx \left(\frac{Kr}{D}\right)^2\int_{0}^{\infty} d\tilde\omega~ \tilde g(\tilde\omega)
\frac{D^2{\tilde\omega}^2}{D^2+{\tilde\omega}^2}\sim r^2\sim(K-K_c)~,
\end{align}
where we used $r\sim(K-K_c)^{1/2}$ in Eq.~\eqref{eq:r_sc}.
Whereas $\delta\dot{Q}=0$ for $K<K_c$, the thermodynamic cost of beating motion for cilia population is reduced as $\delta\dot{Q}\sim (K-K_c)$ upon synchronization for $K\gtrsim K_c$.

\section{Conclusions}
As a simple model to study the generic features of synchronization,
the noisy Kuramoto model is particularly suited to understand the basic thermodynamics involving the synchronization of interacting cilia.
To be specific, the noisy Kuramoto model enabled us to dissect the dissipation from beating cilia into two sources for the case of $K\gtrsim K_c$:
(i) Dissipation, $\omega\omega_s$, arises from the driving force $\omega$ characterizing each cilium's natural beating dynamics in isolation.
(ii) Another dissipation $\dot{\theta}\partial_{\theta}E(\phi,\theta)$ stems from the hydrodynamic coupling between cilia which creates a simple \emph{time-dependent} potential $E(\phi,\theta(t))=-Kr\cos{[\phi-\theta(t)]}$ when the original set of coupled equations is cast into the single-cilium equation at the mean-field level (Eq.~\eqref{eq:Sakaguchi}).
The phase variable of each cilium, $\phi_i$, is attracted towards the average phase angle of the cilia population moving with $\theta(t)=\omega_mt$. The average rate of the heat dissipation from this particular force is called as the Jarzynski contribution.
The total dissipation rate is the sum of the two contributions, leading to Eq.~\eqref{eq:w_i_av}, which is shown to be always non-negative.
We, however, note that quantities different from Eq.~\eqref{eq:w_i_av} have historically been evaluated in the name of viscous dissipation \cite{Taylor1951RoyalSoc,gueron1999PNAS}.
To evaluate dissipation in consistent with the 2nd law of thermodynamics, one has to make sure to include both the contributions (i) and (ii), and particularly (ii), discussed above.

Our model predicts that for the cilia whose beating frequency $\omega$ is slower than the group frequency $\omega_m$ ($|\omega|<|\omega_m|$),
they dissipate more energy upon synchronization ($\langle\dot{q}\rangle_>$) than in isolation ($\langle\dot{q}\rangle_<$), satisfying $\langle\dot{q}\rangle_>  > \langle\dot{q}\rangle_<$.
In fact, this is one of the most interesting points of the present study, as it is seemingly at odds with the general conclusion of reduced dissipation of the whole population upon synchronization, namely, $\langle\dot{Q}\rangle_> \le \langle\dot{Q}\rangle_<$. Despite the presence of such a subpopulation, it is straightforward to prove that dissipation from the whole cilia population always compensates such contribution (see Eq.~\eqref{eq:dQ}).

It has recently been argued by Zhang \emph{et al.} \cite{Zhang2020NatPhys} that in addition to the energy dissipation for driving each individual oscillator, extra energetic cost is required for the oscillator-oscillator coupling in modeling coupled molecular biochemical oscillators, e.g., KaiABC system in the cyanobacterial circadian clock.
They showed that the system is synchronized when the energy dissipation is increased; however, this is in apparent contradiction to the conclusion reached by our noisy Kuramoto model as well as by others \cite{Cao2015NatPhys,pinto2017EPL,lee2018PRE} that the total dissipation from the system is reduced upon synchronization.
The microscopic underpinnings of many-body synchronization may vary from one system to another.
As far as the meso-scale synchronization of beating cilia in a low Reynolds number environment is concerned, there are many experimental evidences that lend support to hydrodynamic interactions as the mechanism of the inter-ciliary coupling and synchronization \cite{goldstein2009PRL,sanchez2011Science,brumley2014Elife,riedel2005Science}.

Finally, all the results of the present study are the logical outcome deduced from a mean field version of the noisy Kuramoto oscillator model. In the biophysical context, however, the emergence of metachronal traveling waves, characterized with both spatial and temporal orderings, would be a more relevant problem to be explored in details, which demands careful considerations of local hydrodynamic couplings, finite size effects, and more realistic natural frequency distributions.

\begin{acknowledgments}
This study was supported by the NRF Grant 2018R1A2B6001790 (HH), 2017R1D1A1B06035497 (HP), "Research Base Construction Fund Support Program" funded by Jeonbuk National University in 2020 (HH), and the KIAS individual Grants PG013604 (HP) and CG035003 (CH) at Korea Institute for Advanced Study.
\end{acknowledgments}

\appendix

\section{normalization constant $Z$}

The normalization constant $Z$ in Eq.~\eqref{eq:pdf_s} is
\begin{align}
Z=\int_0^{2\pi} d\tilde\phi~ {e^{-V(\tilde\phi,\tilde\omega)/D}}
\Bigg(1-\frac{1-e^{-2\pi\tilde\omega/D}}{\int_0^{2\pi}d\phi'e^{V(\phi',\tilde\omega)/D}}\int_0^{\tilde\phi}d\phi''e^{V(\phi'',\tilde\omega)/D}\Bigg)~,
\end{align}
with $V(\tilde\phi,\tilde\omega) \equiv -\tilde\omega\tilde\phi-Kr\cos(\tilde\phi)$. After some algebra using the series expansion
\begin{equation}
e^{-a \cos{x}} = I_0(a) + 2\sum_{n=1}^{\infty} (-1)^n I_n(a)\cos (nx),
\end{equation}
with $I_n(a)$ the $n$-th order modified Bessel function of the first kind, we can easily find
\begin{align}
Z={2\pi}
\frac{I_0^2(a) + 2\sum_{n=1}^{\infty}
 {b^2(-1)^n  I_n^2(a)}/(n^2 + b^2)}
{I_0(a) + 2\sum_{n=1}^{\infty}
 {b^2(-1)^n  I_n(a)}/(n^2 + b^2)}~,\label{eq:app_z}
\end{align}
with $a\equiv Kr/D$ and $b\equiv \tilde\omega/D$.


\section{coupling-modified frequency}

The coupling-modified frequency defined in Eq.~\eqref{eq:omega_s} is
\begin{align}
\tilde\omega_s &=\langle\dot{\tilde\phi}\rangle =
\int_0^{2\pi} d\tilde\phi~ \left[\tilde\omega-Kr\sin {\tilde\phi}\right]~\rho(\tilde\phi,\tilde\omega)~,\nonumber\\
&=\int_0^{2\pi} d\tilde\phi~ \left[-\frac{\partial V}{\partial\tilde\phi}\right]~\rho(\tilde\phi,\tilde\omega)~,\nonumber\\
&=\int_0^{2\pi} d\tilde\phi~\left[  D\frac{\partial\rho}{\partial\tilde\phi}+
\frac{D(1-e^{-2\pi\tilde\omega/D})}{Z\int_0^{2\pi}d\phi'e^{V(\phi',\tilde\omega)/D}}\right]~,\nonumber\\
&=\frac{2\pi D(1-e^{-2\pi\tilde\omega/D})}{Z\int_0^{2\pi}d\phi'e^{V(\phi',\tilde\omega)/D}}~,
\end{align}
where we used the explicit form of $\rho(\tilde\phi,\tilde\omega)$ in Eq.\eqref{eq:pdf_s} and its periodic property $\rho(\tilde\phi,\tilde\omega)=\rho(\tilde\phi+2\pi,\tilde\omega)$.
With the explicit expression of $Z$ in Eq.~\eqref{eq:app_z}, we find
\begin{align}
\tilde\omega_s=\tilde\omega \left[I_0^2(a) + 2\sum_{n=1}^{\infty}
 \frac{b^2(-1)^n  I_n^2(a)}{n^2 + b^2}\right]^{-1}~.
\end{align}

%

\end{document}